\documentclass[prl,letterpaper,aps,10pt,superscriptaddress,twocolumn,floatfix,showpacs]{revtex4-1}
\usepackage{graphicx}
\usepackage{amsmath}
\usepackage{amsfonts}
\usepackage{float}
\usepackage{amssymb}
\usepackage{epsfig}
\usepackage{epstopdf}
\DeclareGraphicsExtensions{.pdf,.eps,.png,.jpg,.mps}
\usepackage[pdftex]{color}
\usepackage{amsmath,graphicx,amssymb,braket,xcolor,subfigure,upgreek}
\usepackage[colorlinks, linkcolor=blue, citecolor=blue, urlcolor=blue, breaklinks=true]{hyperref}
\usepackage{microtype}
\usepackage{bbm}
\usepackage{color}
\usepackage{dsfont}

\begin{document}

\title{Theory of phase-adaptive parametric cooling}
\author{Alekhya Ghosh}
\affiliation{Max Planck Institute for the Science of Light, Staudtstra{\ss}e 2,
D-91058 Erlangen, Germany}
\affiliation{Department of Physics, Friedrich-Alexander-Universit\"{a}t Erlangen-N\"urnberg, Staudtstra{\ss}e 7,
D-91058 Erlangen, Germany}
\author{Pardeep Kumar}
\affiliation{Max Planck Institute for the Science of Light, Staudtstra{\ss}e 2,
D-91058 Erlangen, Germany}
\author{Fidel Jimenez}
\affiliation{Pontificia Universidad Católica del Perú, Av. Universitaria 1801, San Miguel 15088, Peru}
\author{Vivishek Sudhir}
\affiliation{LIGO Laboratory, Massachusetts Institute of Technology, Cambridge, Massachusetts 02139, USA}
\affiliation{Department of Mechanical Engineering, Massachusetts Institute of Technology, Cambridge, MA 02139, USA}
\author{Claudiu Genes}
\affiliation{Max Planck Institute for the Science of Light, Staudtstra{\ss}e 2,
D-91058 Erlangen, Germany}
\affiliation{Department of Physics, Friedrich-Alexander-Universit\"{a}t Erlangen-N\"urnberg, Staudtstra{\ss}e 7,
D-91058 Erlangen, Germany}
\date{\today}

\begin{abstract}
We propose an adaptive phase technique for the parametric cooling of mechanical resonances. This involves the detection of the mechanical quadratures, followed by a sequence of periodic controllable adjustments of the phase of a parametric modulation. The technique allows the preparation of the quantum ground state with an exponential loss of thermal energy, similarly to the case of cold-damping or cavity self-cooling. Analytical derivations are presented for the cooling rate and final occupancies both in the classical and quantum regimes.
\end{abstract}


\maketitle
Mechanical resonators are oftentimes used in displacement~\cite{anetsberger2009displacement}, force~\cite{fogliano2021force}, acceleration~\cite{krause2012acceleration} or mass sensing~\cite{sansa2020mass} applications. Bringing them to their quantum ground state is of fundamental interest for research studying the classical to quantum physics transition~\cite{romero2011quantum}. A particularly successful path in achieving control over motion has been taken in optomechanics where photons are interfaced with phonons~\cite{aspelmeyer2014rmp}. Cooling close to the quantum ground state of an isolated mechanical resonance has been achieved both via cavity self-cooling~\cite{aspelmeyer2019coherent,uros2020cooling,windey2019,gonzalez2019cavitycooling} and via externally imposed feedback~\cite{li2011millikelvin,millen2020review,rossi2018measurement,tebbenjohanns2021quantum,magrini2021real,whittle2021approaching}. In the former case, external driving of a cavity optomechanical system leads to the possibility of resolved sideband cooling~\cite{schliesser2006self,kippenberg2007review}. In the latter case, detection of the light scattered from an optomechanical system allows for the implementation of a cold-damping mechanism~\cite{dania2021feedbackcooling}, where a viscous force proportional to the momentum is provided. Extensions to the simultaneous feedback cooling of many mechanical resonances are also possible~\cite{sommer2019partial,sommer2020multimode}. Both these two paths have been extensively utilized theoretically and experimentally and their advantages and limitations are well understood~\cite{genes2008ground,aspelmeyer2014rmp,komori2022quantum}. More recently, a third option, dubbed parametric cooling, has arisen, as an efficient option for cooling optically levitated particles~\cite{gieseler2012subkelvin,gonzalez2021review}, atoms in cavities~\cite{sames2018continuous} or nano-electromechanical resonators~\cite{villanueva2011a}. For a generic oscillator at resonance frequency $\Omega$, the main ingredient of this technique is the periodic modulation of the spring constant at $2\Omega$. In the particular case of levitated nanoparticles, this is achieved by feedback control of the trapping potential~\cite{gieseler2013thermal}. The cooling mechanism is then similar to cold-damping but with a non-exponential loss rate of energy~\cite{gieseler2012subkelvin,rodenburg2016quantum}.\\
%%%%%%%%%%%%%%%%%%%%%%%
%%%%%%%%%%%%%%%%%%%%%%%
\begin{figure}[b]
\includegraphics[width=0.85\columnwidth]{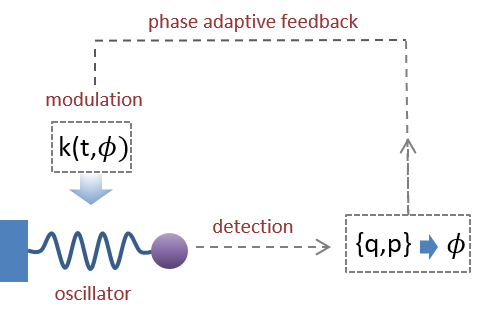}
\caption {\textit{Parametric cooling via phase adaptive feedback.} Generic oscillator with natural frequency $\Omega$. Detection of the mechanical quadratures $q$ and $p$ allows for the design of an adaptive phase feedback strategy, based on the parametric modulation of the spring constant at frequency $2\Omega$ and adjustable modulation phase $\phi$.}
\label{fig1}
\end{figure}
%%%%%%%%%%%%%%%%%%%%%%%
%%%%%%%%%%%%%%%%%%%%%%%
\indent Here, we propose a variation of the parametric cooling technique, which is not based on a controlled cold-damping feedback loop~\cite{tebbenjohanns2021quantum} but, instead, on the adjustment of the modulation phase: we dub this technique \textit{phase adaptive parametric cooling}. The technique solely requires the detection of the momentum and position quadratures of the oscillator, from which an optimal modulation phase is deduced and fed back into the system (as illustrated in Fig.~\ref{fig1}). Parametric driving of an oscillator of displacement $x$ with natural frequency $\Omega$ and mass $m$, consists in supplementing the bare restoring force $\mathcal{F}=-m\Omega^2 x$ with an extra small modulation $\delta \mathcal{F}=2m\Gamma\Omega \cos{(2\Omega t+\phi)}x$ (under the condition $\Gamma\ll\Omega$). Alternatively, this can be understood as a periodic and phase dependent variation of the spring contant $k(t,\phi)$. The key point in our treatment is the observation that the phase $\phi$ is a crucial tuning knob: when properly adjusted, it can steer the system to a cooling regime characterized by an exponential loss of energy at rate $\Gamma$. For a thermal environment at average occupancy $n_\text{th}\gg 1$, we analyze the classical competition between parametric cooling and thermal reheating to predict a final occupancy as low as $\gamma n_\text{th}/\Gamma$, where $\gamma$ is the intrinsic mechanical damping. The same result holds in the quantum regime and can be generalized as well to the simultaneous cooling of a number of oscillators~\cite{brand2021multipleparticles,rudolph2020mutipleparticles,anil2020mutipleparticles}  via a single parametric modulation drive.\\
\indent The emergence of damping behavior is analytically shown by noting that a parametric driven oscillator satisfies a Mathieu-like equation, under perturbative conditions. This analytical solution allows for the identification of the optimal phase of the parametric modulation and the subsequent derivation of an efficient phase adaptive feedback procedure. Classical and quantum limits for cooling efficiency can be derived and compared to simulation of a set of stochastic equations. Finally, the technique can easily be extended to the simultaneous cooling of many mechanical modes.\\

\noindent \textbf{Model and equations} - We consider a mechanical resonance (along some quantized direction $\hat{x}$ of effective mass $m$ subjected to the standard restoring force $\mathcal{F}=-m \Omega^2 \hat{x}$ and to a weak parametric modulation $\delta \mathcal{F}=2 m \Gamma \Omega \cos{(2\Omega t+\phi)} \hat{x}$; here $\Omega$ is the natural oscillation frequency and a modulation amplitude $\Gamma\ll \Omega$ and phase $\phi$ are assumed. With the definition of the zero point motional amplitude $x_\text{zpm}=\sqrt{\hbar/(m\Omega)}$, one can introduce the dimensionless position quadrature $\hat{q}=\hat{x}/x_\text{zpm}$ and the corresponding momentum quadrature $\hat{p}$ such that $[\hat{q},\hat{p}]=i$. The Heisenberg-Langevin equations of motion then can be derived as
\begin{subequations}
\label{Langevin}
  \begin{align}
  \frac{d\hat{q}}{dt} &= \Omega \hat{p},\\
   \frac{d\hat{p}}{dt} &=-\gamma \hat{p}- \Omega \hat{q}+2\Gamma \cos(2\Omega t+\phi)\hat{q}+\hat{\zeta}(t),
  \end{align}
\end{subequations}
describing a parametrically driven, quantum mechanical oscillator subject to thermal noise via the term $\hat{\zeta}(t)$. In the absence of external forces, the mechanical mode is in equilibrium with a thermal bath at temperature $T_\text{th}$ to which an average occupancy $n_\text{th}$ corresponds. The action of the bath onto the mechanical resonance is modeled via quantum stochastic noise operators $\hat{\zeta}(t)$ of zero average and with two-time correlations $\braket{\hat{\zeta}(t)\hat{\zeta}(t')}=1/(2\pi)\textstyle \int d\omega S_\text{th}(\omega) e^{-i\omega (t-t')}$. The thermal power spectrum is given by $S_\text{th}(\omega)=\gamma\omega/\Omega\left\{\coth{[\hbar \omega /(2 k_B T_\text{th})]}+1\right\}$. For large mechanical quality factors $\mathcal{Q}_\text{m}=\Omega/\gamma\gg 1$, the thermal damping rate can be defined in terms of sidebands: $\gamma=[S_\text{th}(\Omega)-S_\text{th}(-\Omega)]/2$ while the average thermal occupancy is $n_\text{th}=[S_\text{th}(\Omega)+S_\text{th}(-\Omega)]/(2\gamma)$. Notice that the equipartition theorem implies that, in a thermal state the variance in the two quadrature is the same and derivable as $\braket{\hat{q}^2}=\braket{\hat{p}^2}=n_\text{th}+1/2$.\\
\indent In a first step, we take an average over the Langevin equations and construct a second order differential equation for the expectation value $q=\braket{\hat{q}}$. We then follow a perturbative method which requires that the modulation parameter $b=\Gamma/\Omega \ll 1$ is small. Eliminating the trivial exponential damping by the transformation $q=\bar{q}e^{-\gamma t/2}$ allows for the derivation of a Mathieu-like equation $\ddot{\bar{q}}+\left[\Omega'^2-2b \Omega^2\cos(2\Omega t+\phi)\right]\bar{q}=0$ where $\Omega'=\sqrt{\Omega^2-\gamma^2/4}$ (as $\mathcal{Q}_\text{m}\gg1$, one can safely approximate this term in the following with $\Omega$). Further simplifications are then obtained via the transformation to a dimensionless time variable $\bar{t}=\Omega t+\phi/2$. The dynamics can now be exactly mapped onto the standard Mathieu equation
  \begin{equation}
  \label{mathieu}
  \frac{d^2 q_M}{d\bar{t}^2}+\left[1-2b \cos(2\bar{t})\right]q_M=0.
  \end{equation}
In the following we will use approximate solution of this equation, obtained in the limit $b\ll1$. Notice that the reverse transformation from $\bar{t}$ to the real time variable is then done via the identification $q(t)=q_\text{M}(\Omega t+\phi/2)e^{-\gamma t/2}$.\\
%%%%%%%%%%%%%%%%%%%%%%%
%%%%%%%%%%%%%%%%%%%%%%%
\begin{figure}[b]
\includegraphics[width=0.95\columnwidth]{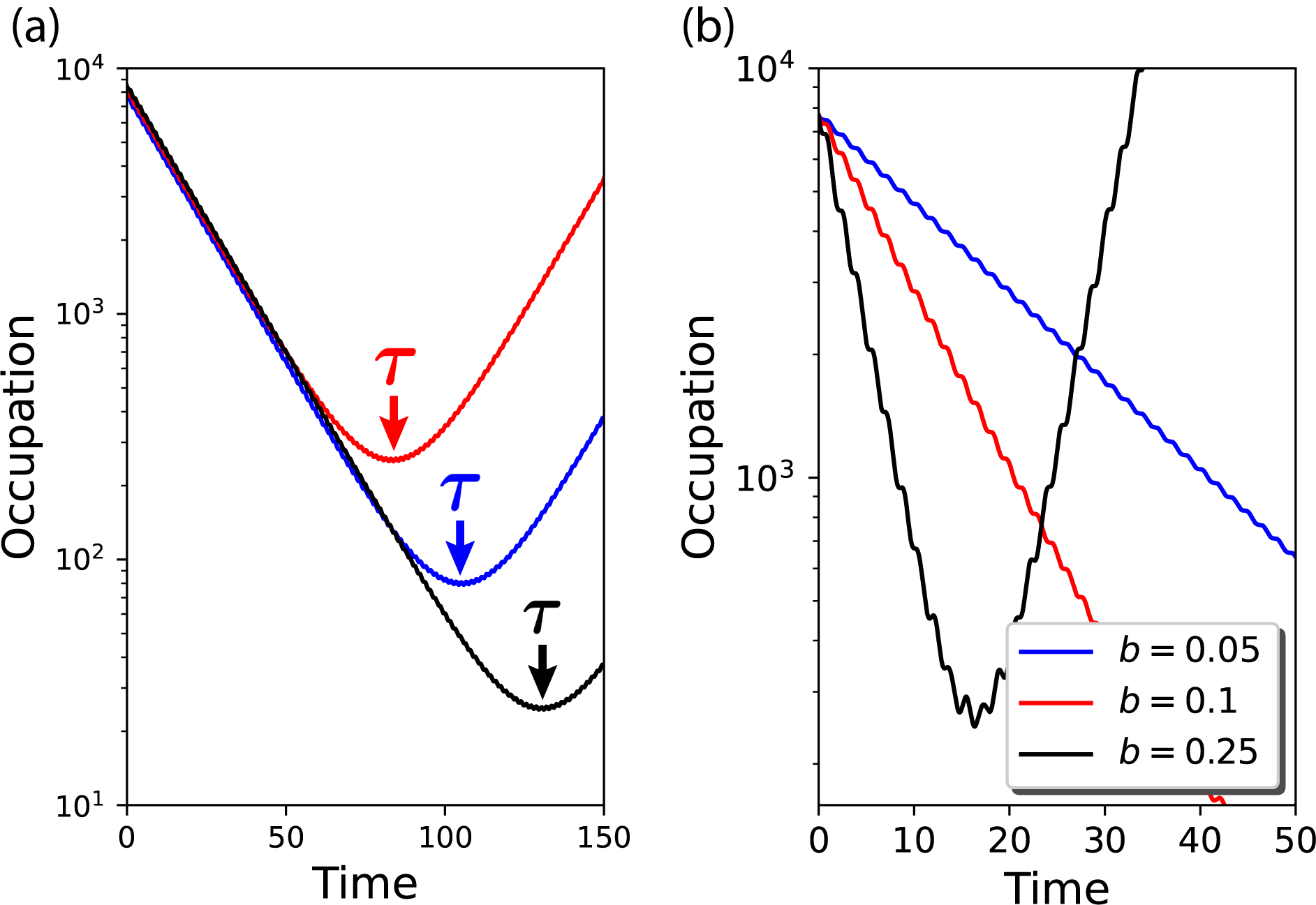}
\caption {\textit{Single shot cooling} a) Time evolution of the occupancy for three randomly selected initial conditions $(q_0^{(j)},p_0^{(j)})$ for $j=1,2,3$, each with an optimized phase $\phi_0^{(j)}$. All trajectories show exponential cooling up to the turning points $\tau_j$ where the damped solution is comparable to the growing one. The parameters are $\gamma/\Omega=10^{-6}$, $n_\text{th}=10^4$ and $b=0.05$. The time is expressed in inverse units of $\Omega$. b) Time dynamics of occupancy for varying $b$ (other parameters same as in (a)) shows the direct dependence of the cooling rate on $b$ and that the validity of the analytically predicted exponential cooling holds even for relatively large $b=0.25$.}
\label{fig2}
\end{figure}
%%%%%%%%%%%%%%%%%%%%%%%
%%%%%%%%%%%%%%%%%%%%%%%
%%%%%%%%%%%%%%%%%%%%%%

\noindent \textbf{Parametric self-cooling solution} -- The general procedure for analytically solving a differential equation in Mathieu form as in Eq.~\eqref{mathieu} is well known~\cite{gradshteyn2007} and famously used, for example, in the theory of harmonic trapping of ions in Paul traps~\cite{leibfried2003quantum} (albeit in a very different regime, where the modulation is much faster than the natural frequency). In the limit of small modulation $b\ll1$, one can show that, to a very good approximation (see Appendix) the (classical) solution can be written in terms of negative and positive damping components
  \begin{align}
  \label{solutions}
  q= \mathcal{A}_-e^{-\frac{(\gamma+\Gamma)t}{2}} \sin{(\Omega t+\phi')}+\mathcal{A}_+e^{-\frac{(\gamma-\Gamma)t}{2}} \cos{(\Omega t+\phi')}
  \end{align}
with $\phi'=\phi/2-\pi/4$. The solution above shows that parametric driving brings in an additional optical damping rate $\Gamma$ (we have assumed $b>0)$ that adds to the intrinsic thermal damping $\gamma$. The negative solution indicates increased damping, while the positive solution leads to a heating instability as soon as $\mathcal{\Gamma}$ becomes larger than $\gamma$. The coefficients $\mathcal{A}_\pm$ are derived from the initial conditions $q_0$ and $p_0$ (as detailed in Appendix) and their ratio strongly depends on the driving phase $\phi$. For a fixed modulation phase $\phi$, the initial conditions then dictate whether the system will follow a cooling or heating dynamics in the transient regime; in the long time limit, independently of the initial conditions, the system will blow up as the heating solution will always prevail.\\
\indent Let us ask for the conditions for which $\mathcal{A}_-\gg\mathcal{A}_+$, such that the damped solution dominates at early times. Under the approximation that $b\ll 1$, a simplified analytical expression can be found (see Appendix)
\begin{equation}
\phi_\text{opt}^{(0)}=\frac{\pi}{2}+2\tan^{-1}\left[\frac{q_0}{p_0}\right],
\label{optphase}
\end{equation}
that sets the proper, optimized value for the modulation phase.
This allows one to formulate the strategy for the single shot cooling mechanism: detection of scattered light allows to infer the initial conditions, which in turn are used to optimize the modulation phase. The resulting cooling dynamics occurs for any initial conditions as illustrated in Fig.~\ref{fig2}. The loss of energy is exponential at the analytically predicted rate $\Gamma$ and dominates up to a time $\tau$ where (roughly) $\tau\approx(1/\Gamma)\ln (\mathcal{A}_-/\mathcal{A}_+) $ (where the positive and negative solutions are comparable). This means that the occupancy reached at time $\tau$ is roughly $\mathcal{A}_-/\mathcal{A}_+$ times smaller than the initial one. Analytically one can deduce $\mathcal{A}_-/\mathcal{A}_+ \approx 2(q_0^2-p_0^2)/(bq_0^2)$ (see Appendix) which indicates on average a reduction in thermal energy by a factor of roughly $1/b$.\\
\indent This result already suggests the mechanism to achieve control over the cooling dynamics at arbitrarily long times: at regular times $j\delta\tau$ (with $j=1...\mathcal{N}$ and the repetition time interval $\delta\tau<\tau$), before heating starts to dominate, one detects the instantaneous values of $q(j\delta\tau)$ and $p(j\delta\tau)$ and updates the phases $\phi(j\delta\tau)$ to the values indicated by Eq.~\ref{optphase}. Following this procedure for $\mathcal{N}$ steps suggests the loss of energy roughly by a factor $1/b^\mathcal{N}$. For values of $b$ around $0.1$, in only $6$ feedback steps one can reduce an initial $n_\text{th}=10^4$ average occupancy to well below unity.\\

%%%%%%%%%%%%%%%%%%%%%%%
%%%%%%%%%%%%%%%%%%%%%%%
\begin{figure}[t]
\includegraphics[width=0.95\columnwidth]{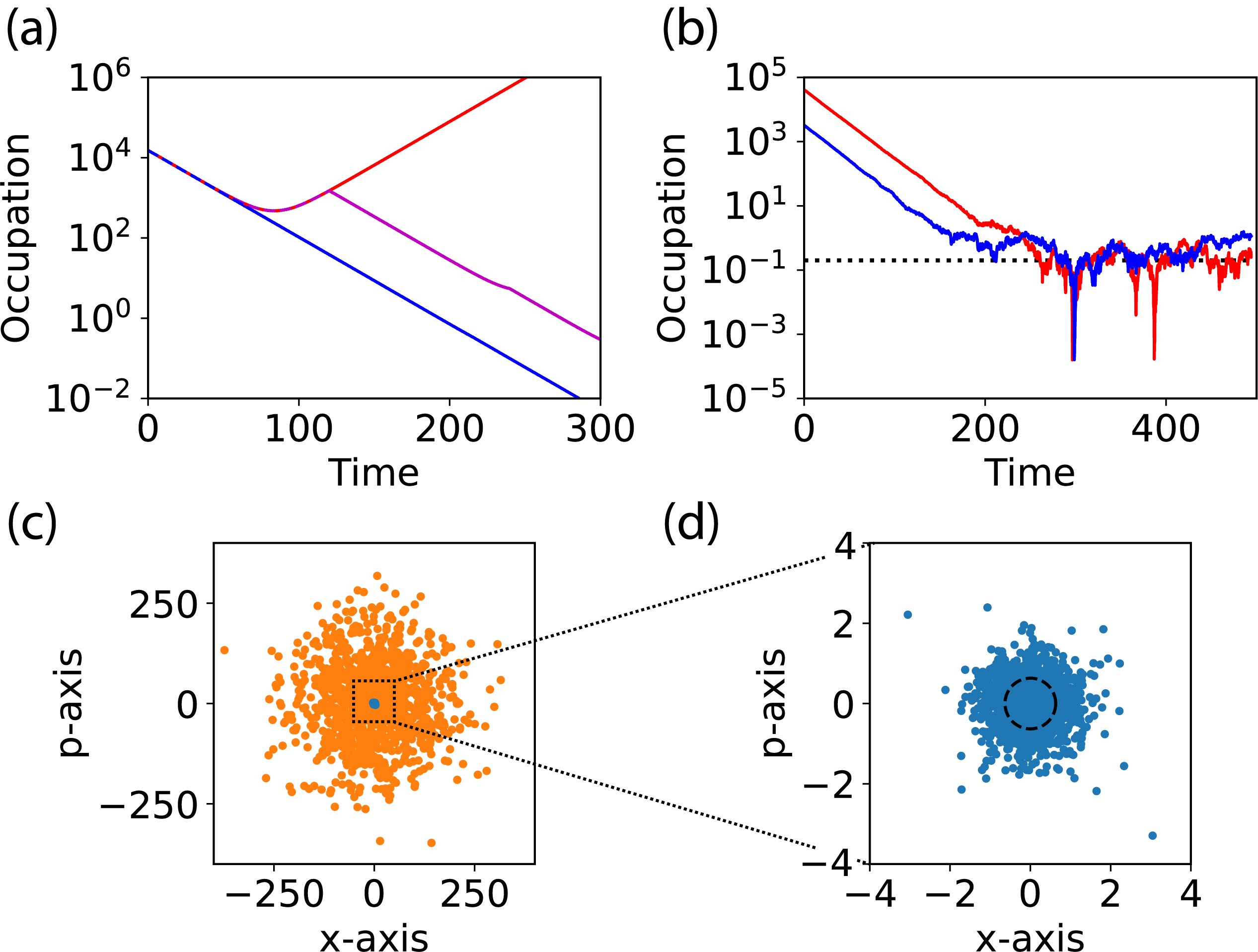}
\caption {\textit{Feedback phase adaptive cooling.} a) Time evolution of the occupancy for a given set of initial conditions assuming initially fixed phase (red line) as opposed to adapted phase with $\delta \tau < \tau$ (blue line, fast feedback) and $\delta \tau>\tau$ (magenta line, slow delayed feedback). b) Two trajectories with different initial conditions randomly picked from an thermal distribution with $n_\text{th}=10^4$. In the presence of thermal noise, the cooling dynamics is purely exponential at the theoretically predicted loss rate $\Gamma$. Close to the ground state, the competition between Brownian noise thermal heating and parametric cooling shows the reach of an equilibrium final occupancy at $n_\text{final}\approx\gamma n_\text{th}/\Gamma$. Fluctuations around this value are owed to the competition between stochastic processes involving thermal reheating and continuous optical cooling action. c) Phase-space illustration of an initial thermal state ($10^3$ points in orange) and the corresponding final cold state (in blue). d) Zoom in of the final occupancy distribution in phase space, showing that the system reaches a thermal state. The parameters are chosen as $\gamma/\Omega=10^{-6}$ and $b=0.05$. The time is expressed in inverse units of $\Omega$. }
\label{fig3}
\end{figure}
%%%%%%%%%%%%%%%%%%%%%%%
%%%%%%%%%%%%%%%%%%%%%%%
%%%%%%%%%%%%%%%%%%%%%%
\noindent \textbf{Feedback cooling versus reheating} -- The performance of any cooling technique involves the competition between the externally induced damping dynamics and the inherent reheating due to the thermal environment in which the oscillator is embedded. We first analyze the performance of an adaptive phase feedback procedure in the presence of classical thermal noise (under the Markovian approximation) modelled by as a Wiener process $\braket{dW(t)}$ of zero average and variance $\braket{dW(t)^2}=dt$ equal to the infinitesimal time increment. To this end one writes a set of coupled difference equations
\begin{subequations}
  \begin{align}
  \label{Wiener}
  dq &= \Omega p dt,\\
  dp &=-\gamma p dt- \Omega q dt+\mathcal{F}_\text{mod}dt+\sqrt{2\gamma n_\text{th}}dW(t),
  \end{align}
\end{subequations}
where the Brownian noise is normalized such that, in the absence of the trap modulation force, the system thermalizes at rate $\gamma$ to the temperature $T_\text{th}$ associated with an average quanta occupancy $n_\text{th}$. For numerical simulations, one can model $dW(t)=\sqrt{dt}\mathcal{N}(0,1)$, where $\mathcal{N}(0,1)$ describes a normally distributed random variable of unit variance. A semi-analytical solution can be instead found by turning the difference equations into a set of recurrence equations. We discretize the time interval $[0,t]$ into $n$ steps of duration $dt = t/n$. The equations above can be rewritten as $\mathbf{v}_n - \mathcal{M}_{n}\mathbf{v}_{n-1} =\sqrt{2\gamma n_\text{th}} \mathbf{u} dW_{n}$ where $\mathbf{v}=(q,p)^\top$ and $\mathbf{u}=(0,1)^\top$ and the evolution matrix is defined as
\begin{equation}
\mathcal{M}_n=\begin{bmatrix}
1 & \Omega dt \\
-\Omega dt+2\Gamma\cos\left[2\Omega(n-1)dt+\phi\right]dt & 1-\gamma dt
\end{bmatrix}.
\end{equation}
A formal analytical solution can be then found easily and can be written in terms of time ordered matrices $\mathcal{T}_{nj}=\mathcal{M}_n\mathcal{M}_{n-1}...\mathcal{M}_j$ in the following form
\begin{equation}
\mathbf{v}_n=\mathcal{T}_{n1} \mathbf{v}_0+\sqrt{2\gamma n_\text{th}}\sum_{j=1}^{n-1}\mathcal{T}_{nj} \mathbf{u} dW_{n-j}.
\end{equation}
The first part in the equation above describes the deterministic evolution from the initial conditions while the last part is the long term behavior dominated by thermal noise. Notice that for $n_\text{th}=0$, the evolution matrix is time independent and the $\mathcal{T}_{nj}$ is simply equal to $\mathcal{M}^{(n-j)}$ which allows for analytical solutions.\\
\indent The phase adaptive feedback algorithm is then as chosen as follows: at the initial time $t=0$, the trapped modulation phase is picked at $\phi_\text{opt}^{(0)}$, from the detected quadratures, to insure an initial damping period. Monitoring of the $q$ and $p$ quadratures at regular time intervals $j\delta\tau$ (with $j=1,...\mathcal{N}$) is then followed by an update of the modulation phases $\phi_\text{opt}^{(j)}$ fixed by Eq.~\eqref{optphase} with the replacement of the instantaneous quadratures $q_j=q(j\delta\tau)$ and $p_j=p(j\delta\tau)$. In Fig.~\ref{fig3}a, the performance of the technique is exemplified on three different trajectories with identical initial conditions but different feedback times (at zero environment temperature). For an optimized $\phi_\text{opt}^{(0)}$ the red line shows the heating in the absence of feedback, while regular interval feedback show either exponential loss of energy when $\delta\tau<\tau$ (blue line) or imperfect exponential loss for $\delta\tau>\tau$ (magenta line). \\
\indent For large thermal occupancies $n_\text{th}=10^4$, the two trajectories (picked randomly from a thermal distribution) shown in Fig.~\ref{fig3}b show that for quick feeedback, exponential loss is achieved towards the same final occupancy. From equilibrium considerations~\cite{genes2008ground}, the final occupancy can be deduced as the ratio of the reheating rate of the ground state $\gamma n_\text{th}$ and the total damping rate $\gamma+\Gamma$, such that $n_\text{final}=\gamma n_\text{th}/(\gamma+\Gamma)$.\\

%%%%%%%%%%%%%%%%%%%%%%%%%%%%%%%%%%%%%%%%%%%%%%%%%
%%%%%%%%%%%%%%%%%%%%%%%%%%%%%%%%%%%%%%%%%%%%%%%%%
%%%%%%%%%%%%%%%%%%%%%%%%%%%%%%%%%%%%%%%%%%%%%%%%%
\noindent \textbf{Quantum ground state cooling} -- The classical model of thermalization does not include the characteristics of the quantum harmonic oscillator, i.e. that the commutations between $\hat{q}$ and $\hat{p}$ which automatically impose a minimal variance of $1/2$ in both quadratures even for zero temperature. To this end, we instead consider the quantum dynamics of the system described by Eqs.~\eqref{Langevin} in differential form and with a quantum white noise input that includes the zero-point energy. Let us assume dynamics around the steady state, characterized by vanishing average momentum and position and by an optimally adjusted modulation phase $\phi$ that gives rise to a constant cooling rate $\Gamma$. We can perform a Fourier transform of Eqs.~\eqref{Langevin} to derive
\begin{equation}
  \label{Fourier}
  \chi^{-1}(\omega) \hat{q} (\omega) =\Gamma \left[e^{i\phi}\hat{q}(-2\Omega+\omega)+ e^{-i\phi}\hat{q}(2\Omega+\omega)\right]+\hat{\zeta}(\omega),
\end{equation}
%%%%%%%%%%%%%%%%%%%%%%%
%%%%%%%%%%%%%%%%%%%%%%%
\begin{figure}[t]
\includegraphics[width=0.95\columnwidth]{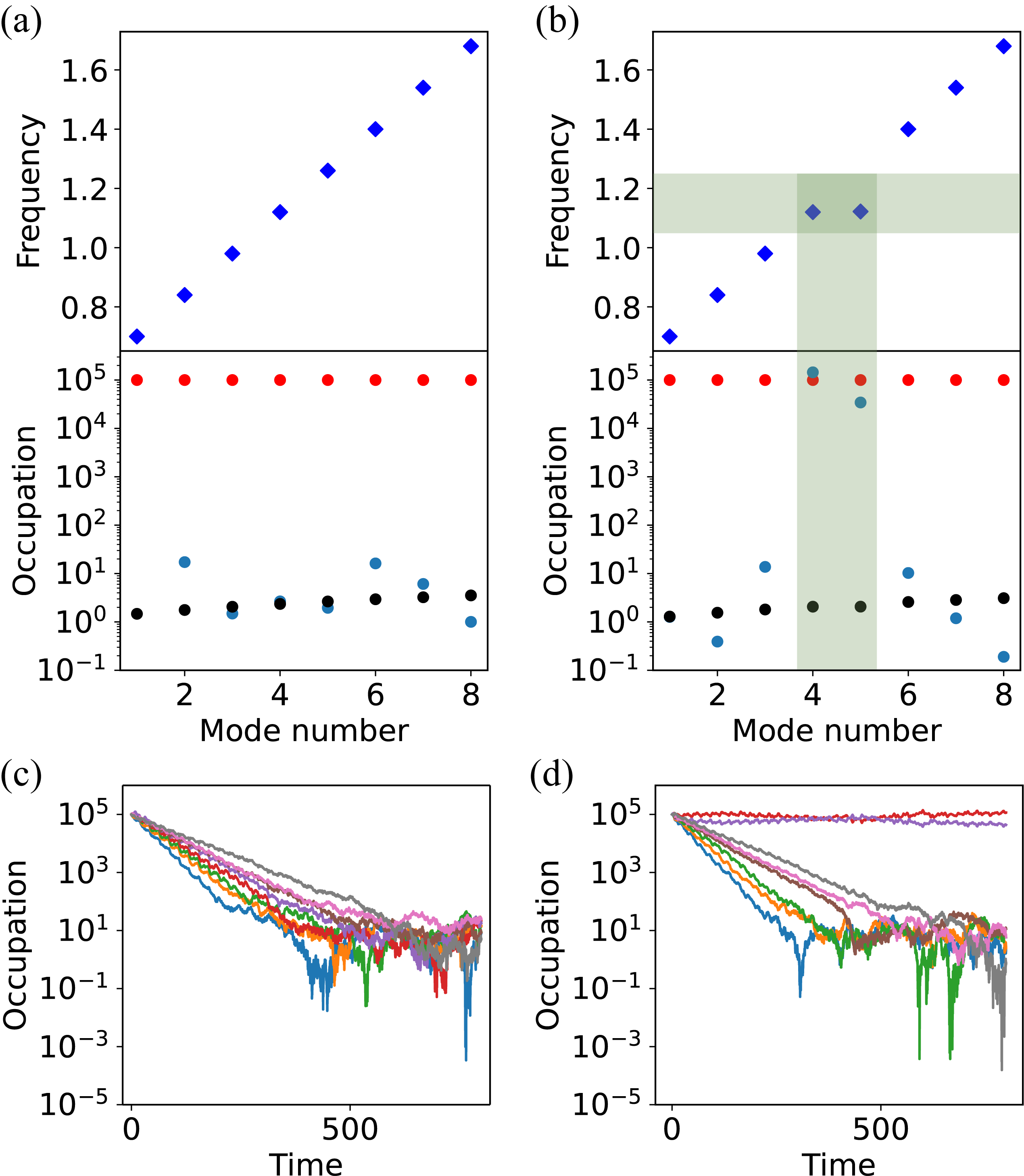}
\caption {\textit{Simultaneous cooling.} a) Final occupancy (lower panel, red - initial occupancy, black - isolated cooling, blue - simultaneous cooling) of eight equidistant modes (upper panel). b) Same as in a) but considering the case where two modes are degenerate in frequency. c) Time dynamics of the cooling process for all modes showing fluctuations associated with the competition between the imposed cooling and the inherent heating dynamics owed to the thermal bath. d) Same as in c) but with the clear message that degenerate modes are decoupled from the cooling dynamics. The parameters are chosen as $\gamma/\Omega=10^{-6}$, $b=0.05$ and the initial occupancy is close to $10^5$ for all modes. The time is expressed in inverse units of $\Omega$.  }
\label{fig4}
\end{figure}
%%%%%%%%%%%%%%%%%%%%%%%
%%%%%%%%%%%%%%%%%%%%%%%
%%%%%%%%%%%%%%%%%%%%%%%%%%%%%%
where the mechanical susceptibility is defined as $\chi(\omega)=\Omega\left(\Omega^2-\omega^2-i \gamma \omega\right)^{-1}$ and the last term describes a sum of noise stemming from both the thermal bath and the effective parametric cooling bath (introduced phenomenologically to account for the fluctuation-dissipation theorem). Similar to the procedure illustrated in Ref.~\cite{genes2008ground}, we compute the power spectrum of the position quadrature and integrate it over all frequencies to obtain the final occupancy as $n_\text{th}=\braket{\hat{q}^2}+1/2$ (as we have assumed that thermal equipartition requires $\braket{\hat{q}^2}=\braket{\hat{p}^2}$). The power spectrum contains Lorentzians of linewidth $\gamma+\Gamma$. Under the assumption that $\gamma+\Gamma\ll \Omega$, one can estimate the final occupancy by only considering power spectra components around $\pm\Omega$. To this end we write $\omega=\pm\Omega+\delta$ where $\delta$ where is a small variation $\delta\ll\Omega$. After integration and under the assumption that $\Gamma>\gamma$ one can show that (see Appendix)
\begin{align}
\braket{\hat{q}^2} = \frac{\gamma(\Gamma+\gamma \sin{\phi})}{\Gamma^2-\gamma^2}\left(n_\text{th}+\frac{1}{2}\right)+ \frac{\Gamma(\Gamma+\gamma\sin{\phi})}{2(\Gamma^2 - \gamma^2)}.
\end{align}
The first contribution describes the effect of the reduction of the thermal bath effect onto the oscillator, while the second term is necessary to secure a larger than $1/2$ variance stemming from quantum fluctuations. A more concise expression can be obtained under the more stringent conditions where $\Gamma\gg \gamma$ and final occupancy $n_\text{final}=\braket{\hat{q}^2}-1/2$ is expressed in a very simple form as
\begin{align}
n_\text{final} =\frac{\gamma}{\Gamma}\left(n_\text{th}+\frac{1}{2}\right)
\end{align}
The result is not surprising and agrees with the stochastic classical analysis previously described: the final temperature is roughly reduced by a factor $\Gamma/\gamma$. This reduction is bounded from above by the mechanical quality factor $\mathcal{Q}_\text{m}$ (as $\Gamma<\Omega$ is the required condition for the validity of the perturbative solution).\\

%%%%%%%%%%%%%%%%%%%%%%%%%%%%%%%%%%%%%%%%%%%%%%%%%
%%%%%%%%%%%%%%%%%%%%%%%%%%%%%%%%%%%%%%%%%%%%%%%%%
%%%%%%%%%%%%%%%%%%%%%%%%%%%%%%%%%%%%%%%%%%%%%%%%%
\noindent \textbf{Simultaneous cooling of multiple modes} -- The phase adaptive mechanism can be extended to simultaneously cool a number $n_\text{res}$ of adjacent mechanical resonances \cite{brand2021multipleparticles,rudolph2020mutipleparticles,anil2020mutipleparticles}. Denoting their frequencies with $\Omega_j$ with $j=1,...n_\text{res}$, we assume a generalized modulation force $\sum_j2\Gamma_j \cos(2\Omega_j t+\phi_j)q_j$. Assuming spectrally resolved detection of all quadratures, the set of modulation phases can be extracted and used for the periodic adjustment of the modulation force at intervals $j \delta \tau$. The results are presented in Fig.~\ref{fig4} both for equidistant, well separated $n_\text{res}=8$ resonances, as well as for the particular case of two degenerate modes. The final occupancy in steady state is calculated for individual, isolated cooling (ignoring the presence of adjacent resonances) and compared with the performance of the simultaneous cooling technique. The results are consistent with previous treatments of cold-damping~\cite{sommer2019partial}, for example, showing that cooling is efficient as long as the modes are frequency separated by more than the effective cooling rate $\Gamma_j$.\\

%%%%%%%%%%%%%%%%%%%%%%%%%%%%%%
%%%%%%%%%%%%%%%%%%%%%%%%%%%%%%%%%%%%%%%%%%%%%%%%%
%%%%%%%%%%%%%%%%%%%%%%%%%%%%%%%%%%%%%%%%%%%%%%%%%
%%%%%%%%%%%%%%%%%%%%%%%%%%%%%%%%%%%%%%%%%%%%%%%%%
\noindent \textbf{Discussions and outlook} --  As opposed to previous theoretical treatments~\cite{rodenburg2016quantum} of parametric cooling \cite{gieseler2012subkelvin}, the technique introduced here does not need external engineering of a cold-damping feedback force and only makes use of a minimally invasive detection loop in the process. The predicted dynamics is purely exponential in energy loss and is captured in a fully analytical model both at the classical and at the quantum level. While parametric cooling has been experimentally introduced for the motional control of optically levitated particles~\cite{gieseler2012subkelvin,gonzalez2021review}, atoms in cavities~\cite{sames2018continuous} or nano-electromechanical resonators~\cite{villanueva2011a}, our description here is quite general involving solely the modulation of the spring constant of the resonator. Therefore, possible applications could extend to optomechanical systems such as a membrane-in-the-middle optomechanical systems, to cooling of ions in traps or to the refrigeration of phonon modes in solid-state systems.\\

%%%%%%%%%%%%%%%%%%%%%%%%%%%%%%%%%%%%%%%%%%%%%%%%%
%%%%%%%%%%%%%%%%%%%%%%%%%%%%%%%%%%%%%%%%%%%%%%%%%
%%%%%%%%%%%%%%%%%%%%%%%%%%%%%%%%%%%%%%%%%%%%%%%%%
\noindent \textbf{Acknowledgments} --  We acknowledge financial support from the Max Planck Society and the Deutsche Forschungsgemeinschaft (DFG, German Research Foundation) -- Project-ID 429529648 -- TRR 306 QuCoLiMa
(``Quantum Cooperativity of Light and Matter'').

\bibliography{Refs}

\onecolumngrid

\appendix

\section{Appendix A: Classical dynamics of a parametrically drive mechanical resonator }
\label{A}

The standard solution to the Mathieu equation is written as an infinite sum
\begin{equation}
q_M(\bar{t})  = \mathcal{E}_{-} e^{i \beta \bar{t}} \sum_{n = -\infty}^{n =\infty} C_{2n} e^{i 2n \bar{t}} + \mathcal{E}_{+} e^{-i \beta \bar{t}} \sum_{n = -\infty}^{n =\infty} C_{2n} e^{-i 2n \bar{t}}.
\end{equation}
The coefficients of all harmonics (see Ref.~\cite{leibfried2003quantum}) can be found to satisfy the following recursive equation
\begin{equation}
C_{2n+2} - D_{2n}C_{2n} + C_{2n-2} = 0,
\end{equation}
where
\begin{equation}
D_{2n} = \frac{1 - (2n + \beta)^2}{b}
\end{equation}
In the limit of small $b$, the first three terms in the expansion above suffice to properly describe the trajectory: we reduce the analysis to $n = 0, \pm1$. For $\beta=-1+x$, where $x$ is a complex number with amplitude much smaller than unity, we find that $D_0=2x/b$ and $D_2=-2x/b$ while all other $D$s are very large. Fixing (without loss of generality) $C_0=1$ and truncating all coefficients for $|n|>1$ we obtain $x=ib$ and $C_2=i$, as the only non-vanishing coefficient. Finally, we can write the solution as
\begin{align}
  q(t) = 2\mathcal{A}_-e^{-\frac{(\gamma+\Gamma)t}{2}}\cos{\left(\Omega t+\phi'\right)}+ 2\mathcal{A}_+e^{-\frac{(\gamma-\Gamma)t}{2}}\sin{\left(\Omega t+\phi'\right)},
\end{align}
where $\mathcal{A}_-=2\mathcal{E}_- e^{i\pi/4}e^{-b\phi/4}$ and $\mathcal{A}_+=2\mathcal{E}_+ e^{i\pi/4}e^{b\phi/4}$ and the newly introduced phase is  $\phi'=\phi/2+\pi/4$. The initial conditions ask that $q_M(\bar{t} = \frac{\phi}{2}) = q(t = 0) = q_0$ and $\dot{q}_M(\bar{t} = \frac{\phi}{2}) = p(t = 0) = p_0$, which leads to the following set of equations
\begin{subequations}
  \begin{align}
  \mathcal{A}_- \cos{\phi'} + \mathcal{A}_+ \sin{\phi'} &= q_0,\\
   -\mathcal{A}_- \sin{\phi'} + \mathcal{A}_+ \cos{\phi'} -\frac{b}{2}\mathcal{A}_- \cos{\phi'}+\frac{b}{2}\mathcal{A}_+\sin{\phi'}&= p_0.
  \end{align}
\end{subequations}
In a first step, we neglect $\mathcal{A}_+$ assuming it is much smaller than the damped solution and obtain from above
\begin{equation}
\phi_\text{opt}^{(0)}=\frac{\pi}{2}+2\tan^{-1}\left[\frac{1}{\frac{p_0}{q_0}+\frac{b}{2}}\right].
\label{phase}
\end{equation}
With the observation that, for the chosen optimal phase, the signs of the $\sin$ and $\cos$ terms are always the same, we can obtain $|\mathcal{A}_-|=\sqrt{q_0^2+p_0^2}$ thus equal to the initial variance of the thermal state.\\
%%%%%%%%%%%%%%%%%%%%%%%%%%%%%%%%%%%%%%%%%%%%%%%%%%%%%%%%%%%%%%%%%%%%%%%%%%%%%%%%%%%%%%%%%%
%%%%%%%%%%%%%%%%%%%%%%%%%%%%%%%%%%%%%%%%%%%%%%%%%%%%%%%%%%%%%%%%%%%%%%%%%%%%%%%%%%%%%%%%%%
\section{Appendix B: Classical stochastic evolution}
\label{B}
%%%%%%%%%%%%%%%%%%%%%%%%%%%%%%%%%%%%%%%%%%%%%%%%%%%%%%%%%%%%%%%%%%%%%%%%%%%%%%%%%%%%%%%%%%
%%%%%%%%%%%%%%%%%%%%%%%%%%%%%%%%%%%%%%%%%%%%%%%%%%%%%%%%%%%%%%%%%%%%%%%%%%%%%%%%%%%%%%%%%%
The coupled difference equations
\begin{subequations}
  \begin{align}
  \label{Wiener}
  dq &= \Omega p dt,\\
  dp &=-\gamma p dt- \Omega q dt+\mathcal{F}_\text{mod}dt+\sqrt{2\gamma n_\text{th}}dW(t),
  \end{align}
\end{subequations}
can be directly numerically simulated by modelling $dW(t)=\sqrt{dt}\mathcal{N}(0,1)$, where $\mathcal{N}(0,1)$ describes a normally distributed random variable of unit variance. However, more analytical insight can be obtained by turning the difference equations into a set of recurrence equations. We discretize the time interval $[0,t]$ into $n$ steps of duration $dt = t/n$ and can then rewrite the equations above as
\begin{equation}
\mathbf{v}_n - \mathcal{M}_{n}\mathbf{v}_{n-1} =\sqrt{2\gamma n_\text{th}} \mathbf{u} dW_{n}
\end{equation}
where $\mathbf{v}=(q,p)^\top$ and $\mathbf{u}=(0,1)^\top$ and the evolution matrix is defined as
\begin{equation}
\mathcal{M}_n=\begin{bmatrix}
1 & \Omega dt \\
-\Omega dt+2\Gamma\cos\left[2\Omega(n-1)dt+\phi\right]dt & 1-\gamma dt
\end{bmatrix}.
\end{equation}
A solution can be then found easily and can be written in terms of time ordered matrices $\mathcal{T}_{nj}=\mathcal{M}_n\mathcal{M}_{n-1}...\mathcal{M}_j$ in the following form
\begin{equation}
\mathbf{v}_n=\mathcal{T}_{n1} \mathbf{v}_0+\sqrt{2\gamma n_\text{th}}\sum_{j=1}^{n-1}\mathcal{T}_{nj} \mathbf{u} dW_{n-j}.
\end{equation}
As a simple check, let us describe solely the thermalization dynamics of an unmodulated oscillator (setting $\Gamma=0$). The time ordered matrices are much simpler now: $\mathcal{T}_{n1}=\mathcal{M}^n$ and $\mathcal{T}_{nj}=\mathcal{M}^{n-j}$. Under the assumption that $\gamma\ll\Omega$, diagonalization of the matrix $\mathcal{M}=S\Lambda S^{-1}$ is straightforward in terms of the two eigenvalues of $\mathcal{M}$ equal to  $\lambda_1=1-\gamma dt/2-i\Omega dt,\, \lambda_2=1-\gamma dt/2+i\Omega dt$. Notice that the two eigenvalues can be rewritten as $\lambda_1=(1-\gamma dt/2)e^{-i\Omega dt}=r e^{i\theta}$ and $\lambda_2=r e^{-i\theta}$ where $r = 1-\gamma t/2n $ and $\theta =\Omega t/n$. The resulting quadratures after $n$ steps are written as
\begin{subequations}
  \begin{align}
  \label{qn-pn}
  q_n &= \frac{\lambda_1^n+\lambda_2^n}{2}q_0+i\,\frac{\lambda_1^n-\lambda_2^n}{2}p_0+\sqrt{2\gamma n_\text{th}}\,\sum_{j=0}^{n-1} \frac{\lambda_1^j-\lambda_2^j}{2}dW_j ,\\
  p_n &=\frac{\lambda_1^n+\lambda_2^n}{2}p_0-i\,\frac{\lambda_1^n-\lambda_2^n}{2}q_0+\sqrt{2\gamma n_\text{th}}\,\sum_{j=0}^{n-1} \frac{\lambda_1^j+\lambda_2^j}{2}dW_j,
  \end{align}
\end{subequations}
where the deterministic parts describe simply the oscillatory weakly damped transient evolution and the last terms are the effect of the thermal environment. In the large $n$ limit we find a closed expression
\begin{subequations}
  \begin{align}
  \label{q-p}
  q(t) &= e^{-\gamma t/2}\left[q_o\cos(\Omega t)+p_0\sin(\Omega t)\right]+\lim_{n\to\infty}\sqrt{2\gamma n_\text{th}}\,\sum_{j=0}^{n-1} r^j\sin(j\theta)dW_j ,\\
  p(t) &=e^{-\gamma t/2}\left[p_0\cos(\Omega t)-q_0\sin(\Omega t)\right]+\lim_{n\to\infty}\sqrt{2\gamma n_\text{th}}\,\sum_{j=0}^{n-1} r^j\cos(j\theta)dW_j,
  \end{align}
\end{subequations}
where we have used that $\lim_{n\to\infty}(1-\gamma t/2n)^n=e^{-\gamma t/2}$. From these expressions, one can estimate that in steady state $\braket{q^2}_\text{ss}=\braket{p^2}_\text{ss}=n_\text{th}$ by using $\braket{dW_j dW_{j'}}=\delta_{jj'}t/n$ and evaluating the limit $\lim_{n\to\infty}\sum_{j=0}^{n-1} r^{2j}\sin^2(j\theta)=1/(2\gamma t)$.\\

\section{Appendix C: Quantum ground state cooling}
\label{C}
Let us assume steady state and fix the modulation phase to $\phi$ such that the cooling rate is exponential at rate $\Gamma$ and both the position and momentum expectation values vanish. We will compute the final occupancy from the variance in position (assuming thermal equipartition between momentum and position quadratures) obtained from the power spectrum in the Fourier domain $S_q (\omega)$ via an integration $\braket{\hat{q}^2} = 1/(2\pi) \int_{-\infty}^{\infty}d\omega S_q (\omega)$. The Fourier transform of Eqs.~\eqref{Langevin} leads to
\begin{subequations}
\label{Fourier}
  \begin{align}
  -i \omega \hat{q}(\omega) &= \Omega \hat{p}(\omega),\\
    -i \omega \hat{p}(\omega) &=-\gamma \hat{p}(\omega)- \Omega \hat{q}(\omega)+\Gamma\left[e^{i\phi}\hat{q}(\omega-2\Omega)+e^{-i\phi}\hat{q}(\omega+2\Omega)\right]+\hat{\zeta}(\omega).
  \end{align}
\end{subequations}
Notice that, for $\Gamma=0$, the above equations lead to $\hat{q} (\omega) =\chi(\omega)\hat{\zeta}(\omega)$ where the mechanical susceptibility is
\begin{equation}
\chi(\omega)=\frac{\Omega}{(\Omega^2-\omega^2)-i \gamma \omega}.
\end{equation}
This indicates a simple way of computing the variance by using the correlations of the thermal environment $\braket{ \zeta(\omega) \zeta(\omega')}= S_\text{th}(\omega) \delta (\omega + \omega')$ where $S_\text{th}(\omega)$ is the thermal power spectrum with sidebands $S_\text{th}(-\Omega)=2\gamma n_\text{th}$ and $S_\text{th}(\Omega)=2\gamma (n_\text{th}+1)$. As for very high mechanical quality factors $\mathcal{Q}_\text{m}\gg 1$, the susceptibility is a very sharply peaked function around $\pm\Omega$ one can approximate it around the two poles at $\pm \Omega$ by expanding it in terms of a small quantity $\gamma\ll|\Delta|\ll\Omega$. The expansion is then $\chi(-\Omega+\Delta)\simeq1/(2\Delta+i\gamma)$ and $\chi(+\Omega+\Delta)\simeq 1/(-2\Delta-i\gamma)$. The variance can then approximated by $\braket{\hat{q}^2}(t) = 1/(2\pi) \int_{-\infty}^{\infty}d\Delta (4\Delta^2+\gamma^2)^{-1} (S_\text{th}(-\Omega)+S_\text{th}(\Omega))=n_\text{th}+1/2$ as expected. We made use of the integral  $1/(2\pi) \int_{-\infty}^{\infty}d\Delta 1/(4\Delta^2+\gamma^2) =\pi/(2\gamma)$.\\
\indent We can now rewrite Eqs.~\eqref{Fourier} as a recursive equation
\begin{equation}
   \hat{q} (\omega) =\Gamma \chi(\omega)\left[e^{i\phi}\hat{q}(-2\Omega+\omega)+ e^{-i\phi}\hat{q}(2\Omega+\omega)\right]+\chi(\omega)\hat{\zeta}(\omega).
\end{equation}
We the proceed as above making small variations around $\pm\Omega$ and assuming that the only components which contribute to the power spectrum of the position quadrature are $\hat{q}(\pm \Omega+\Delta)$. We can now separate two coupled equations
\begin{subequations}
  \begin{align}
 \hat{q}(-\Omega+\Delta) &= \Gamma \chi(-\Omega+\Delta)e^{-i\phi}\hat{q}(\Omega+\Delta)+\chi(-\Omega+\Delta)\hat{\zeta}(-\Omega+\Delta),\\
 \hat{q}(\Omega+\Delta) &= \Gamma \chi(\Omega+\Delta)e^{i\phi}\hat{q}(-\Omega+\Delta)+\chi(\Omega+\Delta)\hat{\zeta}(\Omega+\Delta).
  \end{align}
\end{subequations}
and invert them to find the solutions for $q(-\Omega + \Delta)$ and $q(\Omega + \Delta)$ expressed as
\begin{subequations}
\begin{eqnarray}
 \hat{q}(-\Omega + \Delta) &= -\bar{\chi}(\Delta)\left[ -(2 \Delta + i \gamma) \zeta(-\Omega + \Delta) + \Gamma e^{-i \phi} \zeta(\Omega + \Delta) \right],\\
 \hat{q}(\Omega + \Delta) &= -\bar{\chi}(\Delta)\left[ \Gamma e^{i \phi} \zeta(-\Omega + \Delta) + (2 \Delta + i \gamma) \zeta(\Omega + \Delta) \right].
\end{eqnarray}
\end{subequations}
The modified mechanical susceptibility is approximated by the following expression (under the assumption that $|\Delta| \ll \Omega$)
\begin{equation}
\bar{\chi}(\Delta) = \frac{1}{4 \Delta^2 + \Gamma^2 - \gamma^2 + 4 i \gamma \Delta}.
\end{equation}
The denominator shows the presence of the optical damping rate $\Gamma$ but not as a broadening adding to $\gamma$ as is the case for cold damping or cavity cooling. We now again use the correlations of the noise term in the frequency domain however supplemented with the bath responsible with damping at rate $\Gamma$ and with zero temperature. Notice that, for small $\Delta$, the only contributing terms are $\braket{ \zeta(-\Omega+\Delta) \zeta(\Omega+\Delta')}\simeq S(-\Omega) \delta (\Delta +\Delta')$ and $\braket{ \zeta(\Omega+\Delta) \zeta(-\Omega+\Delta')}\simeq S(\Omega) \delta (\Delta +\Delta')$ where now $S(-\Omega)=2\gamma n_\text{th}$ and $S(\Omega)=2\gamma (n_\text{th}+1)+2\Gamma$. This leads to the following contributions
\begin{subequations}
\begin{eqnarray}
\braket{\hat{q}(-\Omega + \Delta) \hat{q}(-\Omega + \Delta^{'})} & = & |\bar{\chi}(\Delta)|^2 \Gamma e^{-i\phi} \left[2\Delta \left(S(\Omega)-S(-\Omega)\right) - i \gamma\left(S(-\Omega)+S(\Omega)\right)\right] \delta(\Delta + \Delta^{'}),\\
\braket{ \hat{q}(\Omega + \Delta) \hat{q}(\Omega + \Delta^{'})} & = & |\bar{\chi}(\Delta)|^2 \Gamma e^{i\phi} \left[2\Delta \left(S(\Omega)-S(-\Omega)\right) + i \gamma\left(S(-\Omega)+S(\Omega)\right)\right] \delta(\Delta + \Delta^{'}),\\
\braket{ \hat{q}(-\Omega + \Delta) \hat{q}(\Omega + \Delta^{'})} & =  & |\bar{\chi}(\Delta)|^2 \left[(4\Delta^2+\gamma^2)S(-\Omega)+\Gamma^2 S(\Omega))\right] \delta(\Delta + \Delta^{'}),\\
\braket{ \hat{q}(\Omega + \Delta) \hat{q}(-\Omega + \Delta^{'})} & = & |\bar{\chi}(\Delta)|^2 \left[\Gamma^2 S(-\Omega)+(4\Delta^2+\gamma^2) S(\Omega))\right] \delta(\Delta + \Delta^{'}).
\end{eqnarray}
\end{subequations}
We can now add all the contributions to find the position power spectrum via the following integral
\begin{equation}
\braket{\hat{q}^2} = \frac{1}{2\pi} \int_{-\infty}^{\infty}d\Delta |\bar{\chi}(\Delta)|^2 \left\{ 4\gamma(2\gamma\Gamma\sin{\phi}+4\Delta^2+\Gamma^2+\gamma^2)(n_\text{th}+\frac{1}{2})+2\Gamma\left(4\Delta (\gamma+\Gamma) \cos{\phi}+2\gamma \Gamma \sin{\phi}+4\Delta^2+\Gamma^2+\gamma^2\right)\right\}
\end{equation}
The result for the relevant case $\Gamma>\gamma$ is
\begin{align}
\braket{\hat{q}^2} = \frac{\gamma(\Gamma+\gamma \sin{\phi})}{\Gamma^2-\gamma^2}\left(n_\text{th}+\frac{1}{2}\right)+ \frac{\Gamma(\Gamma+\gamma\sin{\phi})}{2(\Gamma^2 - \gamma^2)}.
\end{align}
Notice that in the limit where the additional damping dominates $\Gamma\gg \gamma$ one can simplify the expression above and compute the final occupancy $n_\text{final}=\braket{\hat{q}^2}-1/2$ to lead to
\begin{align}
n_\text{final} =\frac{\gamma}{\Gamma}(n_\text{th}+\frac{1}{2})
\end{align}
In the opposite case where $\Gamma<\gamma$ the variance reads
\begin{align}
\braket{\hat{q}^2} = \frac{\gamma(\gamma+\Gamma \sin{\phi})}{\gamma^2-\Gamma^2}\left(n_\text{th}+\frac{1}{2}\right)+ \frac{\Gamma(\gamma+\Gamma\sin{\phi})}{2(\gamma^2 - \Gamma^2)}.
\end{align}
For $\Gamma=0$ we recover the expected result $\braket{\hat{q}^2}=n_\text{th}+1/2$.
\end{document}